%% file: bare_jrnl_new_sample4.tex
\documentclass[lettersize,journal]{IEEEtran}
\usepackage[nolist]{acronym}
\usepackage{amsmath,amssymb,amsfonts}
\usepackage{array}
\usepackage{booktabs}
\usepackage{cite}
\usepackage{graphicx}
\usepackage{xfrac}

\DeclareMathOperator{\sinc}{sinc}

\begin{document}
\input{acronym}

\title{Frequency Selective Reflection of Wideband Signals with Reconfigurable Intelligent Surfaces}

\author{Pedro~H.~C.~de~Souza and Luciano~Mendes
\thanks{Pedro H. C. de Souza, and Luciano Mendes, are with the National Institute of Telecommunications - Inatel, Sta. Rita do Sapucaí-MG, Brazil (e-mails: pedro.carneiro@dtel.inatel.br, luciano@inatel.br). This work has been funded by the following research projects: Brasil 6G Project with support from RNP/MCTI (Grant 01245.010604/2020-14), xGMobile Project code XGM-AFCCT-2024-2-15-1 with resources from EMBRAPII/MCTI (Grant 052/2023 PPI IoT/Manufatura 4.0) and FAPEMIG (Grant PPE-00124-23), SEMEAR Project supported by FAPESP (Grant No. 22/09319-9), SAMURAI Project supported by FAPESP (Grant 20/05127-2), Ciência por Elas with resources from FAPEMIG (Grant APQ-04523-23), Fomento à Internacionalização das ICTMGs with resources from FAPEMIG (Grant APQ-05305-23), Programa de Apoio a Instalações Multiusuários with resources from FAPEMIG (Grant APQ-01558-24), and Redes Estruturantes, de Pesquisa Científica ou de Desenvolvimento Tecnológico with resources from FAPEMIG (Grant RED-00194-23). This work has also been supported by a fellowship from CNPq and FAPESP.}
}

\markboth{Journal of \LaTeX\ Class Files,~Vol.~14, No.~8, August~2021}%
{Shell \MakeLowercase{\textit{et al.}}: A Sample Article Using IEEEtran.cls for IEEE Journals}


\maketitle

\begin{abstract}
Recently, the \ac{RIS} technology has ushered in the prospect of control over the wireless propagation environment. By establishing alternative propagation paths for the transmitted signals, and by reflecting them in a controllable manner, the \ac{RIS} is able to improve the signal reception. However, an aspect often overlooked is the potential bandwidth restrictions on the wideband signal reflected by the \ac{RIS}. If not carefully considered, this can become an impediment for the adoption of the \ac{RIS} in the next generation of communications systems. Therefore, in this work we propose a \ac{RIS} configuration method that provides frequency selective signal reflection for wideband signals. 
\end{abstract}


\section{Introduction}
\acresetall

\IEEEPARstart{W}{hile} the \ac{5G} is being deployed, and well underway globally \cite{eric:25,nperf}, the shaping of the next generation of mobile communications also begins to acquire more definite traces. Study cases and requirements for the \ac{6G} are now being discussed by the academia and industry alike and several new technologies are under scrutiny. More specifically, the \ac{RIS} technology is currently on track to be considered as a study-item by important standardization bodies, such as the \ac{3GPP} \cite{astr:24}.

In general, the \ac{RIS} can be described as a planar surface composed by multiple elements or reflectors, each of them built with tunable resonant elements. These passive reflectors provide adjustable impedance values \cite{liaskos:22}, which can then be used to rotate (phase-shift) the electromagnetic waves impinging on its surface. As long as the \ac{RIS} is properly positioned between a transmitter and receiver, it is able to reflect signals through additional propagation paths \cite{bjorn:22}. Consequently, the \ac{RIS} can create favorable propagation conditions for the signal transmission, specially if the direct channel between the transmitter and receiver is in \ac{NLOS}.

Typically, the \ac{RIS} reflectors are configured in such a way as to promote the most constructive combination of signals at the receiver \cite{pan:22}. This ensures that the channel capacity at the receiver is increased, since the signal reflected by the \ac{RIS} is now being combined coherently with signal received via the direct channel. However, the interference management of the signals reflected by the \ac{RIS} is one aspect that is overlooked by the aforementioned approach \cite{astr:24}. On the other hand, works such as \cite{mueller:25}, for example, propose a \ac{RIS} hardware design that provides a greater degree of control over the reflected signals, allowing a frequency selective \ac{RIS} configuration to be attained. Furthermore, in \cite{sena:25}, an evolution of the \ac{RIS} technology, known as \ac{BD-RIS}, is used for multi-band operation in a \ac{MIMO} communications system. In \cite{kastanos:24}, an algorithm is proposed for configuring the precoding vectors and the \acp{RIS}' phase configurations in a distributed manner, taking into account the frequency selectivity behavior of the \ac{RIS}.

Nevertheless, in this work we propose a time-variant \ac{RIS} configuration that is capable of selecting sub-bands for wideband signal reflection. More specifically, we demonstrate that a frequency selective signal reflection can be achieved within a bandwidth of interest, and without special requirements for the \ac{RIS} hardware. To the best of authors' knowledge, such a frequency selective \ac{RIS} configuration method for wideband signal reflection is, as of yet, unexplored for interference management in \ac{RIS}-aided communication systems.

This work is organized as follows: Section~\ref{sec:sysmodel} introduces the signal propagation model supporting the proposed \ac{RIS}-aided communications system; Section~\ref{sec:RISconfig} details the \ac{RIS} configuration method that achieves a frequency selective signal reflection; in Section~\ref{sec:numresult} the main findings are discussed with the assistance of numerical results generated via computational simulations and, finally, Section~\ref{sec:conclusion} concludes the paper.

\subsection{Notation}
Throughout this paper, the $n$th entry of the vector $\mathbf{x}$ is represented by $x\left[n\right]$. The entry on the $i$th row and $j$th column of the matrix $\mathbf{X}$ is denoted by $X_{i,j}$. The number of elements contained in a set $\mathcal{X}$ is denoted by $\left\lvert\mathcal{X}\right\rvert$. The rounding of scalars to the nearest integer is represented by $\lfloor x\rceil$, whereas $\lfloor x\rfloor$ rounds to the nearest integer smaller than $x \in \mathbb{R}$. The sets of vectors (matrices) of dimension $X$ ($X\times Y$) with real and complex entries are respectively represented by $\mathbb{R}^{X}$ ($\mathbb{R}^{X\times Y}$) and $\mathbb{C}^{X}$ ($\mathbb{C}^{X\times Y}$). The transposition and conjugate transposition operations of a vector or matrix are represented as $\left(\cdot\right)^{\text{T}}$ and $\left(\cdot\right)^{\text{H}}$, respectively. The $X$-dimensional vector of all ones is denoted by $\mathbf{1}_{X}$. The $\ell_p$-norm, $p \geq 1$, of the vector $\mathbf{x} \in \mathbb{C}^{N}$ is given by $\|\mathbf{x}\|_p$ and also $\left\langle\mathbf{x}\right\rangle = N^{-1}\sum_n x\left[n\right]$ for $\mathbf{x} \in \mathbb{R}^{N}$. Finally, $x \sim \mathcal{U} \left[a\text{,}b\right]$ is a random value drawn from the uniform distribution over the interval $\left[a,b\right]$ and $x \sim \mathcal{N}\left(0,1\right)$ is a random value drawn from the normal distribution.

\section{System Model}\label{sec:sysmodel}
Let a point-to-point communications system between a transmitter, denoted as an \ac{AP}, and a receiver or \ac{UE}, be assisted by a \ac{RIS} with $N$ reflectors or elements. The precise \ac{RIS} location is conveniently determined so that the \ac{AP} and \ac{UE} have \ac{LOS} propagation paths with the \ac{RIS}. Therefore, the \ac{RIS} can then enhance the signal reception in a scenario where the direct signal propagation paths between the \ac{AP} and \ac{UE} are in \ac{NLOS}. In other words, by reflecting signals transmitted from the \ac{AP} back to the \ac{UE}, the \ac{RIS} creates new propagation paths with more favorable conditions for signal transmission. For this, the $n$-th \ac{RIS} reflector is configured with a continuous phase-shift, $\theta_n = \left[-\pi,\pi\right] \ \forall n \in \{1\text{,}2\text{,}\dots\text{,}N\}$. 

\begin{figure}[h!]
	\centering
	\includegraphics[width=0.95\columnwidth,keepaspectratio]{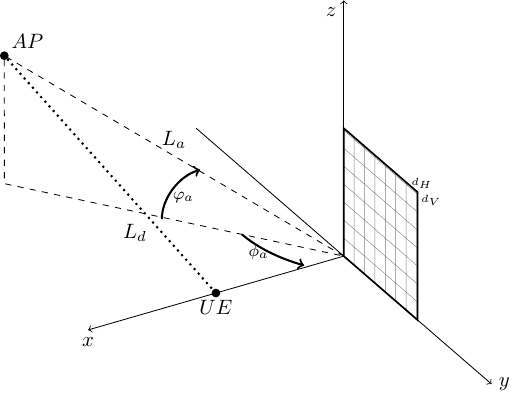}
	\caption{Diagram of the propagation model on a $\left(x\text{, } y\text{, } z\right)$ coordinate system.}\label{fig:sysdiag}
\end{figure}

Figure~\ref{fig:sysdiag} illustrates the communications system discussed in this work. Observe in Figure~\ref{fig:sysdiag} that the direct channel is represented by $L_d$ (\ac{NLOS}) propagation paths between the \ac{AP} and \ac{UE}, whereas the composite channel, composed by the cascade of the channels between the \ac{AP} and \ac{RIS} and \ac{RIS} to the \ac{UE}, have $L_a$ and $L_b$ propagation paths, respectively. The relative azimuth angle of arrival, $\phi_a$, and the elevation angle, $\varphi_a$, are also being illustrated in Figure~\ref{fig:sysdiag}\footnote{The redundant $L_b$, $\phi_b$ and $\varphi_b$ symbols are omitted to improve visibility.}. These angles are used to compute the array response of the \ac{RIS} \cite{pedro:25}, which accounts for different phase rotations depending upon the location of a given reflector across the \ac{RIS} plane. Moreover, note that each reflector has sides of size $d_\text{H}$ and $d_\text{V}$ meters. 

More specifically, the \ac{RIS} array response, considering the $L_a$ channel paths, can be given by
\begin{equation}\label{eq:spcsig}
    \mathcal{S}\left(\Phi_a\right) = e^{\jmath {\boldsymbol{\phi}^{(l)}_a}^\text{T} \Psi}\text{, with } \Psi = [\boldsymbol{0}\text{,}\Psi_\text{H}\text{,}\Psi_\text{V}]^\text{T}\text{,}
\end{equation}
where $l \in \{1\text{,}2\text{,}\dots\text{,}L_a\}$ and furthermore we have
\begin{equation}
    \scalebox{0.79}{$\Phi^{(l)}_a = \frac{-2\pi}{\lambda}\left[\cos{\left(\phi^{(l)}_a\right)}\cos{\left(\varphi^{(l)}_a\right)}\text{,}\sin{\left(\phi^{(l)}_a\right)}\cos{\left(\varphi^{(l)}_a\right)}\text{,}\sin{\left(\varphi^{(l)}_a\right)}\right]^\text{T}\text{,}$}
\end{equation}
in which $\lambda = 3 \times 10^8 f_c^{-1}$, $f_c$ being the central frequency of the signal carrier. Additionally, consider the following
\begin{subequations}
    \begin{align}
        \Psi_\text{H} &= d_\text{H}\left[\bmod{(\sfrac{0}{N_\text{row}})}\text{,}\bmod{(\sfrac{1}{N_\text{row}})}\text{,}\dots\text{,}\bmod{(\sfrac{N-1}{N_\text{row}})}\right]^\text{T}\text{;} \\[.25em]
        \Psi_\text{V} &= d_\text{V}\left[\lfloor\sfrac{0}{N_\text{col}}\rceil \text{,}\lfloor\sfrac{1}{N_\text{col}}\rceil \text{,}\dots\text{,}\lfloor\sfrac{N-1}{N_\text{col}}\rceil\right]^\text{T}\text{,}
    \end{align}
\end{subequations}
for which $N = N_\text{row}N_\text{col}$, where $N_\text{row}$ and $N_\text{col}$ are the number of rows and columns of the \ac{RIS}, respectively. Observe that $\Psi \in \mathbb{C}^{3\times N}$ can be seen as the area spanned by the \ac{RIS} planar surface across the $y-z$ plane, as illustrated in Figure~\ref{fig:sysdiag}. In other words, the \ac{RIS} reflector farthest from the plane origin imposes more rotation to the signal, with the other reflectors presenting gradually less rotation as they draw closer to the plane origin. These rotation values are then weighted by the impinging signal angles (angles of arrival), with all these operations being computed accordingly in \eqref{eq:spcsig}. The same formulation of \eqref{eq:spcsig} is employed for the $L_b$ channel paths, the main difference being that the angles of departure are now considered instead of the angles of arrival. Note also that the \ac{RIS} orientation could be modified without loss of generality.

Consequently, the discrete-time impulse response of the composite channel can be written as
\begin{align}\label{eq:rxsignal}
    h_c\left[m\right] = \sum_{n=0}^{N-1}\sum_{l=1}^{L_a}\sum_{\ell=1}^{L_b}&\sqrt{\alpha^{(l)} \beta^{(\ell)}}\mathcal{S}\left(\Phi_a\right)\mathcal{S}\left(\Phi_b\right) \\
    &\times s\left(m\right) e^{-\jmath 2\pi f_c\left(\tau^{(l)}_a + \tau^{(\ell)}_b + \tau_{\theta_n^t}\right)} \text{;} \nonumber
\end{align}
in which $m \in \{1\text{,}2\text{,}\dots\text{,}M\}$ and also,
\begin{equation}
    s\left(m\right) = \sinc\left(m + B\left(\eta - \tau^{(l)}_a - \tau^{(\ell)}_b + \tau_{\theta_n^t}\right)\right)\text{,}
\end{equation}
where $\alpha^{(l)}$ and $\beta^{(\ell)}$ are, respectively, the propagation losses for the $L_a$ and $L_b$ channels, $\mathcal{S}\left(\Phi_a\right)$ and $\mathcal{S}\left(\Phi_b\right)$ are given by \eqref{eq:spcsig}, $\tau^{(l)}_a \geq 0$ and $\tau^{(\ell)}_b \geq 0$ are the propagation delays, whereas ${\tau_{\theta_n^t} = \theta_n^t / 2\pi f_c}$ is the time-varying phase-shift imposed by the $n$-th \ac{RIS} element at the discrete-time index $t \in \{1\text{,}2\text{,}\dots\text{,}T\}$. Moreover, $B$ is the bandwidth occupied by all subcarriers. Likewise, the direct channel can be given as
\begin{align}\label{eq:signal}
    h_d\left[m\right]  = \sum_{l=1}^{L_d}&\sqrt{\delta^{(l)}} \sinc\left(m + B \left(\eta - \tau^{(l)}_d\right)\right) \\
    &\times e^{-\jmath 2\pi \left(f_c \tau^{(l)}_d\right)} \text{,}
\end{align}
wherein $\delta^{(l)} \ \forall l \in L_d$ are the propagation losses for the $L_d$ direct channel paths and $\tau^{(l)}_d \geq 0$ represent the propagation delays. Note that $\eta = \min{(\tau^{(l)}_d)}$, $\forall l \in L_d$, both in \eqref{eq:signal} and \eqref{eq:rxsignal} in order to assure causality.

In the next section, we show that it is possible to configure the \ac{RIS} in such a way as to obtain frequency selective reflection for wideband signals.

\section{RIS Configuration}\label{sec:RISconfig}
It is also assumed, for the \ac{RIS} assisted communications system presented in this work, that a wideband signal is transmitted by the \ac{AP} using the \ac{OFDM} system. In this scenario, typically the \ac{RIS} is configured with the aim of maximizing the achievable rate at the \ac{UE} \cite{bjorn:22,pedro:25}. More importantly, it is assumed that the signal transmitted on all $K$ \ac{OFDM} subcarriers are to be reflected by the \ac{RIS} to an intended \ac{UE}. However, this may impose restrictions on the \ac{RIS} implementation if only a subset of subcarriers are allocated to this \ac{UE}. In other words, the unrestricted reflection of wideband signals by the \ac{RIS} may become undesirable, since the reflected signal to unintended \acp{UE} can occur if the remaining subcarriers are allocated to them. With that in mind and by assuming a constant channel during the \ac{RIS} configuration, then we can write the achievable rate \cite{bjorn:22,pedro:25} as follows
\begin{equation}\label{eq:rate}
    R = \frac{B}{\xi}\sum_{i=0}^{K - 1}{\log_2{\left(1 + \frac{p_i\|\mathbf{f}_i^\text{H} \mathbf{h}_d\left\lvert\mathcal{I}\right\rvert^{-1} + \mathbf{f}_i^\text{H} \boldsymbol{\vartheta} \|_2^2}{BN_0}\right)}} \ \sfrac{\text{bit}}{\text{s}}\text{.}
\end{equation}
By letting $\mathcal{I}$ denote the set of subcarriers indexes (bins) selected for the selective signal reflection, then we also have $\mathbf{f}_i$ representing the $i$-th row of the \ac{DFT} matrix $F_{i,j} = e^{-\jmath 2\pi ij/K}$, $B$ as the bandwidth occupied by the selected subcarriers, $\xi = \lvert\mathcal{I}\rvert + M - 1$, to take into account the cyclic prefix loss, with $\mathbf{p} \in \mathbb{R}^K$ being the power vector, in which $p_i$ is the power allocated to the $i$-th subcarrier, such that $P = \left\langle \mathbf{p}\right\rangle$; $P$ being the total transmission power, and, finally, $N_0$ as the \ac{AWGN} power density. Moreover, see in \eqref{eq:rate} that $\mathbf{h}_d \in \mathbb{C}^{K}$ denotes $K$ samples (with $K - M$ padding samples) of the discrete-time impulse response of the direct channel as given in \eqref{eq:signal}. The composite channel samples can be otherwise expressed as
\begin{equation}\label{eq:compch}
    \boldsymbol{\vartheta} = \left(\mathbf{V} \boldsymbol{\Omega}_{\theta}\right)^\text{T} \boldsymbol{1}_N\text{,}
\end{equation}
wherein $\mathbf{V} \in \mathbb{C}^{N \times K}$ contains the discrete-time impulse response for all $N$ composite channels and $\boldsymbol{\Omega}_{\theta} \in \mathbb{C}^{K \times T}$ is the time-varying \ac{RIS} configuration. More specifically, the entries of the \ac{RIS} configuration matrix are defined as follows:
\begin{equation}\label{eq:matrixEntries}
    \boldsymbol{\Omega}_{\theta} = \begin{bmatrix}
    \theta_n^1 & \theta_n^2 & \cdots & \theta_n^T \\
    \theta_n^1 & \theta_n^2 & \cdots & \theta_n^T \\
    \vdots & \vdots & \ddots & \vdots \\
    \theta_n^1 & \theta_n^2 & \cdots & \theta_n^T \\
    \end{bmatrix}\text{;}
\end{equation}
which differs from the typical diagonal matrix used to represent the \ac{RIS} configuration. However, observe that $\boldsymbol{\Omega}_{\theta}$ encodes a time-varying \ac{RIS} configuration, for which its columns can be seen as the phase-shift applied to all samples of the composite channel's impulse response at the discrete-time index $t$. In other words, the columns of $\boldsymbol{\Omega}_{\theta}$ represent a given \ac{RIS} configuration at time $t$. For the sake of simplicity, we henceforth drop the subscript $n$ of the phase-shift configurations, such that $\theta^t = \theta_1^t = \theta_2^t = \dots = \theta_n^t$, meaning that all $N$ \ac{RIS} reflectors are configured with the same phase-shift. This will become clear in what follows.

\subsection{Defining the Configuration Matrix $\boldsymbol{\Omega}_{\theta}$}
It is well known that in order to properly estimate channel coefficients in \ac{RIS}-aided communication systems, one may resort to orthogonal matrices \cite{ozdo:20}. Similarly, in order to achieve a frequency selective signal reflection, in this work we define the \ac{RIS} configuration matrix according to
\begin{equation}\label{eq:RISmatrix}
\boldsymbol{\Omega}_{\theta} = \boldsymbol{1}_K \frac{\mathbf{w}^\text{T}}{\left\lvert\mathcal{I}\right\rvert}\text{; where } \mathbf{w} = \mathbf{F}\mathbf{b} \in \mathbb{C}^{K}
\end{equation}
for which $\mathbf{F} \in \mathbb{C}^{K \times K}$ is the \ac{DFT} matrix and $b\left[i\right] = 1$ for $i \in \mathcal{I}$ and zero otherwise. It is easy to show that this \ac{RIS} configuration leads to a frequency selective reflection. Firstly, note that the right hand side (composite channel term) of the euclidean norm of \eqref{eq:rate} can be rewritten as
\begin{subequations}
    \begin{align}
        \mathbf{f}_i^\text{H} \boldsymbol{\vartheta} &=\left\lvert\mathcal{I}\right\rvert^{-1}\mathbf{f}_i^\text{H}\mathbf{w}\sum_{\forall n}{\boldsymbol{1}^\text{T}_K \mathbf{v}_n}\text{,} \label{eq:FSeq_a} \\
        &=\left\lvert\mathcal{I}\right\rvert^{-1}\mathbf{f}_i^\text{H}\mathbf{F}\mathbf{b}\sum_{\forall n}{\sum_{\forall k}{v_n [k]}}\text{,} \label{eq:FSeq_b}
    \end{align}
\end{subequations}
where $\mathbf{v}_n$ represents the $n$-th row of $\mathbf{V}$. Consequently, by knowing that $\mathbf{F}^\text{H}\mathbf{F}\mathbf{b} = \mathbf{b}$, then it follows that
\begin{equation} \label{eq:FSrslt}
    \mathbf{f}_i^\text{H} \boldsymbol{\vartheta} = \left\{ 
    \begin{matrix}
        &\left\lvert\mathcal{I}\right\rvert^{-1}\sum_{\forall n}{\sum_{\forall k}{v_n [k]}}\text{, } &i \in \mathcal{I}\text{,} \\ 
        &0\text{, } &i \notin \mathcal{I}\text{.}
    \end{matrix} 
    \right.
\end{equation}
See that the term $\mathbf{f}_i^\text{H}\mathbf{w}$ in \eqref{eq:FSeq_a} essentially depends on $\mathbf{b}$, which acts as a frequency bin selector based on the subcarriers indexes stored in $\mathcal{I}$. Moreover, note that the columns of the \ac{RIS} configuration matrix (see \eqref{eq:RISmatrix}) are, in turn, determined by the normalized linear combinations of the \ac{DFT} matrix columns. Therefore, the \ac{RIS} reflectors are only required to be configured with a time-varying continuous phase-shift, through which each reflector remains passive, that is, $\|\boldsymbol{\Omega}_{\theta_{i,j}}\|_2^2 \leq 1$ $\forall i\text{, }j$. 

As stated before, the row entries of $\boldsymbol{\Omega}_{\theta}$ are fixed, changing only when a new configuration is needed depending on the time index $t$. Note, however, that $T \geq K$, meaning that the minimum number of required configurations scales with the total number of subcarriers. This is equivalent to say that more time resources are needed as more bandwidth is used for the signal transmission. Therefore, more stringent requirements on the channel coherence time could pose a challenge for the frequency selectivity of the reflected signal, since multiple \ac{RIS} configurations must be performed while the channel remains constant. This will be explored further in Subsection~\ref{subsec:freqS}. 

\subsection{Motivation: A General Communications Scenario}
\begin{figure}[h!]
	\centering
	\includegraphics[width=0.85\columnwidth,keepaspectratio]{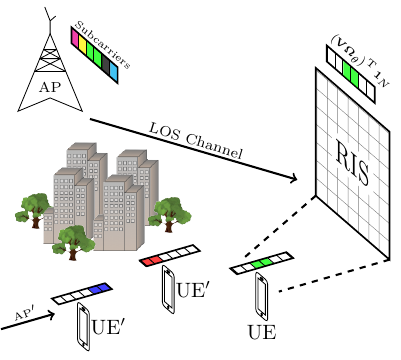}
	\caption{A communications scenario where the \ac{RIS} is employed without interfering with the signal received by other \acp{UE} (i.e. \ac{UE}$^\prime$). Note that the represented dimensions are not to scale and that this figure is better visualized in colors.}\label{fig:FSdiag}
\end{figure}

Figure~\ref{fig:FSdiag} demonstrates a general communications scenario where the frequency selective reflection could be employed. It shows the intended \ac{UE} receiving the \ac{AP} reflected signal only on the selected subcarriers, whereas others \acp{UE} in the vicinity (denoted as \ac{UE}$^\prime$), are not hampered by unrestricted signal reflections. In principle, this means that signals transmitted by other \acp{AP}, that is, \ac{AP}$^\prime$, are not going to be inadvertently interfered by signals reflected from the \ac{RIS}. 

In what follows, computational simulations are leveraged to generate numerical results that corroborates the frequency selective properties of the proposed \ac{RIS} configuration method.

\section{Numerical Results}\label{sec:numresult}
In this work, numerical results are obtained under the following considerations for the communications system presented in Section~\ref{sec:sysmodel}: (i) we ascribe $L_a = 101$ propagation paths for the \ac{AP}-\ac{RIS} channel, with each path delay randomly varying according to $\tau^l_a \sim \mathcal{U} \left[\tau^1_a\text{,}2\tau^1_a\right]\text{, } \forall l > 1$ (the first path, $l = 1$, is the strongest \ac{LOS} path); (ii) the same goes for the \ac{RIS}-\ac{UE} channel with $L_b = 51$ path delays; (iii) all $L_d = 100$ path delays of the direct channel (\ac{AP}-\ac{UE}) are randomly distributed such that $\tau^l_d \sim \mathcal{U} \left[\tau_d\text{,}2\tau_d\right]$ ($\tau_d$ is the time, in seconds, that the signal would take to propagate through the direct channel under \ac{LOS} conditions); (iv) the azimuth and elevation angles of arrival/departure shown in Figure~\ref{fig:sysdiag} can also vary randomly around the \ac{LOS} path initial angle, with $\boldsymbol{\phi}_{a,b} \sim \mathcal{U} \left[-40^{\circ}\text{,}40^{\circ}\right]$ and $\boldsymbol{\varphi}_{a,b} \sim \mathcal{U} \left[-10^{\circ}\text{,}10^{\circ}\right]$; (v) moreover, let $f_c = 3$ GHz be the central frequency of the transmitted signal, $B = 10.5$ MHz and $N_0=-164$ dBm; and, finally, (vi) the dimensions of the \ac{RIS} reflector, as indicated in Figure~\ref{fig:sysdiag}, are defined to be $d_\text{H} = d_\text{V} = 0.25\lambda$ meters. The remaining parameters for the communications system \cite{pedro:25} under analysis are described in Table~\ref{tbl:chparam}.
\input{tables/chparams}

\subsection{Frequency Selectivity Performance}\label{subsec:freqS}
To determine how much of the reflected signal is received by unintended \acp{UE}, that is, the portion of the \ac{RIS} reflection interfering on unselected subcarriers, we propose the following calculation
\begin{equation}\label{eq:SovI}
    \frac{\text{S}}{\text{I}} = \frac{\sum_{\forall i}{\|\mathbf{f}_i^\text{H} \mathbf{h}_d\left\lvert\mathcal{I}\right\rvert^{-1} + \mathbf{f}_i^\text{H} \boldsymbol{\vartheta} \|_2^2}}{\sum_{\forall j}{\|\mathbf{f}_j^\text{H} \boldsymbol{\vartheta} \|_2^2}}\text{,}
\end{equation}
in which $i \in \mathcal{I}$ and, consequently, $j \notin \mathcal{I}$. In other words, the ratio between the combined signal magnitude (`S') of the selected subcarriers, over the combined signal magnitude (`I') of the remaining or interfered subcarriers, give us a metric to evaluate the frequency selectivity of the \ac{RIS} reflection. Note that $\mathbf{f}_j^\text{H}\mathbf{h}_d = 0\text{, } \forall j$, for simplicity, since we only wish to account for the interfering reflected signal on unselected subcarriers. From \eqref{eq:FSrslt} it follows that \eqref{eq:SovI} results in $\text{S}/\text{I} \rightarrow \infty$, granting a perfect selective reflection of signals. However, this result holds if at least $T = K$ is guaranteed. Conversely, if the number of discrete-time \ac{RIS} configurations is insufficient, that is, if $T < K$, it can lead to an incomplete set of \ac{DFT} coefficients being employed by the \ac{RIS} configuration. As a consequence, the interfering reflected signals may become more prominent and less frequency selectivity is achieved. 

\begin{figure}[h!]
	\centering
	\includegraphics[width=0.95\columnwidth,keepaspectratio]{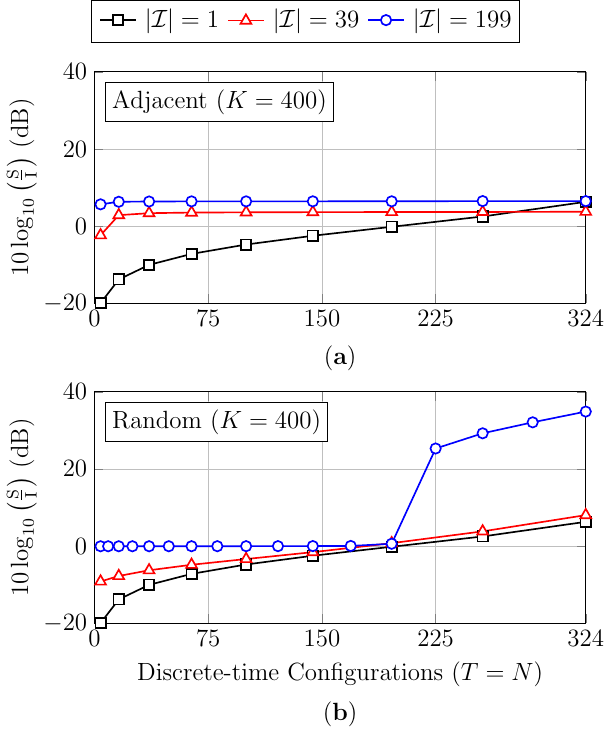}
	\caption{The ratio between the combined signal magnitude of the selected subcarriers, over the signal magnitude of the remaining subcarriers is traced for different numbers of discrete-time configurations (or \ac{RIS} sizes) and selection methods. Scenarios with $\lvert\mathcal{I}\rvert = 1\text{, }39$ and $199$ selected subcarriers are evaluated. Each simulation point consists of averaging (up to) 5$\times 10^3$ channel realizations.}\label{fig:FSSovI}
\end{figure}

In Figure~\ref{fig:FSSovI}, the results of \eqref{eq:SovI} are traced for different \ac{RIS} sizes of $N$ reflectors, where for the sake of brevity $T = N$. It was observed that the frequency selectivity performance depends largely on the number of discrete-time configurations, $T$. Consequently, the results for $N \neq T$ are omitted. Moreover, the total number of subcarriers\footnote{The optimum power allocation is computed by the water filling algorithm.} is fixed to $K = 400$ and the number of selected subcarriers can vary from a single subcarrier $\lvert\mathcal{I}\rvert = 1$ to $\lvert\mathcal{I}\rvert = 39$ and $\lvert\mathcal{I}\rvert = 199$ subcarriers. As can be observed in Figures~\ref{fig:FSSovI} (a) and (b), a pair of different selection methods are also adopted. The adjacent (Figure~\ref{fig:FSSovI} (a)) selection of subcarriers uses a reference subcarrier index, being drawn at random with equal probability, which is then surrounded by an even number of subcarriers for selective signal reflection. Alternatively, the random (Figure~\ref{fig:FSSovI} (b)) selection of subcarriers uses a random set of subcarriers, where subcarriers indexes are also chosen with equal probability, but which are not necessarily adjacent. The set of subcarriers indexes according to these selection methods can be given by
\begin{subequations} \label{eq:idxtarget}
    \begin{align}
        \mathcal{I} &= \left\{j-\left\lfloor\frac{\left\lvert\mathcal{I}\right\rvert}{2}\right\rfloor\text{,}\dots\text{,}j-1\text{,}j\text{,}j+1\text{,}\dots\text{,}j+\left\lfloor\frac{\left\lvert\mathcal{I}\right\rvert}{2}\right\rfloor\right\}\text{,} \label{eq:idxtarget_adj} \\[.5em] 
            \mathcal{I} &\sim \lfloor\mathcal{U} \left[0\text{,}K - \left\lvert\mathcal{I}\right\rvert\right]\rceil\text{,} \label{eq:idxtarget_rnd}
    \end{align}
\end{subequations}
wherein $j = \{0\text{,}1\text{,}\dots\text{,}K-1\}$ and for which we have the adjacent and random selection methods represented in \eqref{eq:idxtarget_adj} and \eqref{eq:idxtarget_rnd}, respectively. Notice that for the random selection, $\lvert\mathcal{I}\rvert$ indexes are drawn with no reposition and that consecutive indexes are not allowed.

Therefore, see in Figure~\ref{fig:FSSovI} (a) that more interference is observed as the number of discrete-time configurations diminishes, specially when a single subcarrier ($\lvert\mathcal{I}\rvert = 1$) is selected for reflection. Likewise, in Figure~\ref{fig:FSSovI} (b), a similar behavior is observed, except that an inflection point is present for $T \sim 200$ when $\lvert\mathcal{I}\rvert = 199$ selected subcarriers are considered. Because the random method does not allow reposition and consecutive subcarriers selection, it becomes practically deterministic in this case. Although the indexes are still chosen at random, it is granted that every other subcarrier is eventually selected when $K/2$ subcarriers are used. In other words, this results in a fixed alternate subcarrier selection (e.g. $\mathcal{I} = \{0\text{,}2\text{,}4\text{,}\dots\text{,}K-2\}$). Consequently, the subcarriers used for selective reflection are evenly spaced across all the bandwidth $B$. As verified in Figure~\ref{fig:FSSovI} (b), this has the effect of making the selective reflection considerable better for $T > 200$.

\subsection{Achievable Rate Performance}
By assuming that the condition for perfect selective reflection is met ($T = K$), then it remains to be investigated what is the achievable rate performance. For this, we leverage the relative rate \cite{pedro:25}, which is defined as follows
\begin{equation}\label{eq:relrate}
    R_\text{Rel.} = \sfrac{R}{R_\text{C}} \times 100 \ \left(\%\right)\text{,}
\end{equation}
wherein $R$ is the achievable rate of \eqref{eq:rate} and the coherent rate, $R_\text{C}$, is defined by the maximum achievable rate. The coherent rate can be seen as the upper bound on the achievable rate, since it is based on the ideal conditions for which the direct and composite channels combine (in-phase) coherently at the \ac{UE} \cite{bjorn:22,pedro:25}. In summary, the relative rate allows us to quantify how reflected signals combine (constructively or destructively) with the signals transmitted via the direct channel. 

\begin{figure}[h!]
	\centering
	\includegraphics[width=0.95\columnwidth,keepaspectratio]{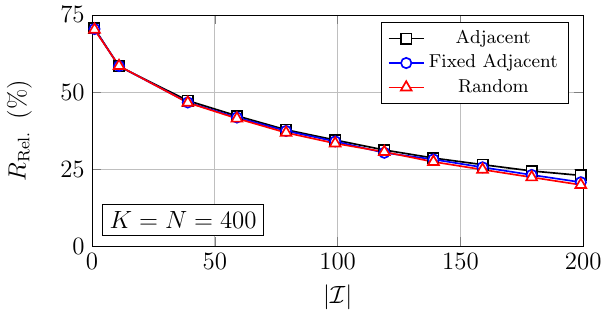}
	\caption{Relative rate results are traced for a range of different quantities of selected subcarriers, also considering that $K = N = 400$. Besides the selection methods of \eqref{eq:idxtarget_adj} and \eqref{eq:idxtarget_rnd}, the fixed adjacent method is also illustrated. Each simulation point consists of averaging (up to) 5$\times 10^3$ channel realizations.}\label{fig:FSrate}
\end{figure}

Figure~\ref{fig:FSrate} shows the relative rate results for a range of different quantities of selected subcarriers, where we also have $K = N = 400$. Additionally, the selection methods of \eqref{eq:idxtarget_adj} (Adjacent) and \eqref{eq:idxtarget_rnd} (Random) are also present, with the addition of the fixed adjacent method. This method includes a restriction on the indexes available for the reference subcarrier, such that $\lfloor\lvert\mathcal{I}\rvert/2\rfloor \leq \min(j\text{,}K-j-1)$ is always true. Consequently, the total number of selected subcarriers is always $\lvert\mathcal{I}\rvert$ for the fixed adjacent method.

Recall that the \ac{RIS} configuration is being used for a frequency selective signal reflection and, as such, its main objective is not necessarily the channel capacity or rate maximization. Nonetheless, note in Figure~\ref{fig:FSrate} that $R_\text{Rel.} \simeq 75$ \% for a low number of selected subcarriers, representing a satisfactory performance. The relative rate performance, however, decreases steadily as more subcarriers are used for selective reflection. This contrasts with the results observed in Figure~\ref{fig:FSSovI}, where generally more selected subcarriers mean better selectivity performance. Therefore, a trade-off exists between achievable rate and selectivity performance when determining the number of selected subcarriers. Observe also in Figure~\ref{fig:FSrate} that the relative rate of the adjacent method is marginally better in relation to the other methods. This is caused by its lower average number of selected subcarriers, which is reassured by the fact (see also \eqref{eq:idxtarget_adj}) that the fixed adjacent has a slightly lower relative rate. 

Furthermore, it was verified that the general results observed in Figure~\ref{fig:FSSovI} (a) are also true for the fixed adjacent method, with slightly higher values of $S/I$ for the fixed adjacent selection (considering $\lvert\mathcal{I}\rvert = 39\text{, }199$). The results obtained with higher values for $K$ and, consequently, $N$ and $T$, have also shown similar behavior to that observed in Figure~\ref{fig:FSSovI} and Figure~\ref{fig:FSrate}, except that the $S/I$ and relative rate values are generally higher. These results are omitted since their findings do not bring any new relevant information.

\section{Conclusions}\label{sec:conclusion}
In this work, a frequency selective \ac{RIS} configuration was put forward. It was shown that this configuration allows the selection of an arbitrary subset of \ac{OFDM} subcarriers for the signal reflection performed by the \ac{RIS}. This configuration method avoids signal reflections in sub-bands that are not necessarily allocated for the intended receiver. Numerical results have also demonstrated that the proposed \ac{RIS} configuration can provide beneficial signal reflection properties and interference management, even under non-ideal conditions. For future works, practical \ac{RIS} hardware limitations and their impact on the reflected signal properties should be also investigated. 
 
\bibliography{references}
\bibliographystyle{IEEEtran}

\end{document}

%% file: acronym.tex
\begin{acronym}
        \acro{3GPP}{3rd generation partnership project}
        \acro{5G}{fifth generation of mobile networks}
        \acro{6G}{sixth generation of mobile networks}
        \acro{AO}{alternating optimization}
	\acro{AP}{access point}
        \acro{AWGN}{additive white Gaussian noise}
        \acro{BD-RIS}{beyond diagonal reconfigurable intelligent surface}
        \acro{DFT}{discrete Fourier transform}
        \acro{LOS}{line-of-sight}
        \acro{MIMO}{multiple-input multiple-output}
        \acro{MISO}{multiple-input single-output}
        \acro{MLP}{multilayer perceptron}
        \acro{NLOS}{non-line-of-sight}
        \acro{NN}{neural network}
        \acro{OFDM}{orthogonal frequency division multiplex}
        \acro{ReLU}{rectified linear unit}
        \acro{RIS}{reconfigurable intelligent surface}
        \acro{SCA}{successive convex approximation}
        \acro{STM}{strongest tap maximization}
        \acro{UE}{user equipment}
\end{acronym}

%% file: tables/chparams.tex
\begin{table}[h!]
	\caption{Fixed parameters of the channel model. Observe that $u \sim \mathcal{U} \left[0\text{,}1\right]$ and $n \sim \mathcal{N}\left(0,1\right)$.}\label{tbl:chparam}
	\centering
	\setlength{\tabcolsep}{5pt}
	\resizebox{0.95\columnwidth}{!}{%
	\begin{tabular}{>{\raggedright\arraybackslash}m{0.25\linewidth}>{\raggedright\arraybackslash}m{0.75\linewidth}} \toprule
    \textbf{Parameter} & \textbf{Value} \\ \midrule
    \multicolumn{2}{c}{\textit{Coordinates}} \\ \midrule
    $\left(x_a\text{, } y_a\text{, } z_a\right)$ & 
    $\left(40\text{, } -200\text{, } 0\right)$ meters \\[1em] 
    $\left(x_b\text{, } y_b\text{, } z_b\right)$ & 
    $\left(20\text{, } 0\text{, } 0\right)$ meters \\[1em]
    \multicolumn{2}{c}{\textit{Number of Samples}} \\ \midrule
    $M$ & 
    $\displaystyle \left\lfloor B \left(2\tau^{(1)}_a + 2\tau^{(1)}_b - \tau^{(1)}_d\right)\right\rceil + 11$ samples \\[2em] 
    \multicolumn{2}{c}{\textit{Propagation Losses}} \\ \midrule
     $\alpha^{(l)}$ (as in $\beta^{(\ell)}$) & 
     $\displaystyle \gamma^{(l)} \frac{d_H d_V}{\lambda^2 / 4\pi} 10^{\frac{-30.18 - 26\log{\left(d_a\right)}}{10}}$\\[3em] 
    $\gamma^{(l)}$ & 
    $\displaystyle \frac{\omega^{(l)}_P}{\sum_{\forall l}{\omega^{(l)}_P}\left(1 + \omega_R\right)} \text{ , } \omega^{(l)}_P = 10^{-\tau^{(l)}_a + 0.2 n} \text{ , } \forall l>1$ \\[3em]
    $\tau^{(1)}_a$ & 
    $\displaystyle \frac{\overset{d_a}{\overbrace{\|\left(x_a\text{, } y_a\text{, } z_a\right)\|_2^2}}}{3 \times 10^8}$ \\[4em]
    $\gamma^{(1)}$ & 
    $\displaystyle \frac{\omega_R}{1 + \omega_R} \text{ , } \omega_R = 10^{\frac{13 - 0.03 d_a}{10}}$ \\
    \bottomrule
    \end{tabular}
    }
\end{table}